\documentclass[12pt]{iopart}

\usepackage{iopams}  
\expandafter\let\csname equation*\endcsname\relax
\expandafter\let\csname endequation*\endcsname\relax
\usepackage{amsmath}
\usepackage{mathtools}
\usepackage{xcolor}
\usepackage{graphicx}
\usepackage{caption}
\usepackage{comment}
\usepackage{bm}
\usepackage{enumerate}
\usepackage[colorlinks=true,linkcolor=blue,citecolor=blue,urlcolor=blue,
hyperfootnotes]{hyperref}
\usepackage{footmisc}
\usepackage{float}
\usepackage{xfrac}
\usepackage{latexsym}
\usepackage{amstext}
\usepackage[T1]{fontenc}
\usepackage{hyperref}
\usepackage{cleveref}
\usepackage{etoolbox}
\begin{document}

\def\snt#1{\textcolor{red}{#1}}
\def\snc#1{\textcolor{red}{\tt [SN: #1]}}
\def\SK#1{\textcolor{magenta}{#1}}

\title[Gravitational Waves in the Circular Restricted Three Body Problem]{Gravitational Waves in the Circular Restricted Three Body Problem}

\author{{Mikel Martin Barandiaran$^{1}$, Sachiko Kuroyanagi$^{1,2}$, Savvas Nesseris$^{1}$}}
\ead{mikel.martin@uam.es, sachiko.kuroyanagi@csic.es, savvas.nesseris@csic.es}
\address{1.- Instituto de F\'isica Te\'orica UAM-CSIC, Universidad Aut\'onoma de Madrid,
Cantoblanco, 28049 Madrid, Spain.}
\address{2.- Department of Physics and Astrophysics, Nagoya University, Nagoya, 464-8602, Japan}

\begin{abstract}
The prospect of unprecedented high-quality data of gravitational waves in the upcoming decades demands a theoretical effort to optimally study and analyze the signals that next generation detectors will provide. Here we study the gravitational wave emission and related dynamics during the inspiralling phase of the Circular Restricted Three Body Problem, a modification of the conventional binary scenario in which a small third object co-rotates with the parent binary system. Specifically, we obtain analytic expressions for the emitted power, frequency variation and other dynamical variables that describe the evolution of the system. As a key highlight, we find that the presence of the third body actually slows down the coalescence of the binary, which can be partially interpreted as an effective rescaling of the binary's chirp-mass. Our analysis assumes semi-Keplerian orbits for the particles and a highly mass asymmetric parent binary needed for the stability of orbits.
\end{abstract}
\begin{indented}
\item[]February 2024 \hfill IFT-UAM/CSIC-23-122
\end{indented}

\section{Introduction}

The field of Cosmology is going through a revolution. Since the first direct detection of gravitational waves (GWs) by the LIGO-Virgo (LV) collaboration in 2015~\cite{firstGW}, a whole new world of possibilities has opened up for the physics community. Already predicted back in 1916~\cite{einstein1,einstein2} and considered one of the major successes of General Relativity (GR, hereafter), the study of GWs unfolds an entirely novel channel for observing the Universe. The field of GW research has gathered enormous momentum within the scientific community, and several missions set to revolutionize our understanding on these ripples in the fabric of space-time are already underway.

To this date, a total of 94 binary coalescence events\footnote{Since the criteria for considering candidate signals as statistically significant detections may vary among authors, this number might slightly change throughout the literature.} have been detected by the LVK collaboration~\cite{94events}. In the future, as we anticipate a broader range of sources combined with an improved sensitivity to GWs, it becomes crucial to establish a solid theoretical framework and compile an exhaustive list of potential GW sources. This paper is dedicated to exploring a modest alteration to the binary scenario, involving the addition of a third small body. 

The three-body problem has drawn attention as a challenge within the domain of general relativity due to its lack of a general analytical solution, with only a partial solution having been accomplished to date. Its application to GW observation has also been extensively discussed in the literature~\cite{Nakamura:1983,Wardell:2002iq, Gultekin:2005fd, Chiba:2006ad, Campanelli:2007ea, Torigoe:2009bw, Asada:2009qs, Galaviz:2010te,  Dmitrasinovic:2014lha}. More recently, it has gained even greater interest due to the growing focus on GW astronomy~\cite{Fiziev:2016mhx, Meiron:2016ipr, Bonetti:2018tpf, Gupta:2019unn, Lim:2020cvm, Will:2020tri, Kuntz:2021hhm, Kuntz:2022juv}.

This paper aims to provide a comprehensive discussion of the prospects for observing GWs in the context of the Circular Restricted Three Body Problem (CR3BP). The CR3BP is a classical problem in celestial mechanics that describes  the motion of a small mass under the gravitational influence of two larger masses, which are in planar circular orbits. In earlier studies, it was shown that Lagrange's triangle solution yields a quadrupole waveform whose functional form is the same as that of the waveform produced by a binary system~\cite{Torigoe:2009bw}. Then, in Ref.~\cite{Asada:2009qs}, the GW waveforms were derived to demonstrate that the octupole part of the waveform allows for differentiation between a binary system and a triple one. We revisit the quadrupole problem putting focus on the case of the Lagrange triangular points L4 and L5, paying particular attention to the condition where the mass of the third body is significantly smaller than that of the primary binary. This mass asymmetry is essential to ensure the stability of the third body's orbit. 

Additionally, we derive the quadrupole waveform in the Fourier domain, which is useful for the template search and compare it with the two body case. Finally, we discuss the orbit's stability through numerical computations. The stability of the triangular Lagrange points is studied earlier in \cite{2010CeMDA.107..145S,Schnittman:2010br,Yamada:2012wm,Huang:2014gor,Yamada:2015epa}.

Our paper is organized as follows: in Sec.~\ref{sec:binarysection} we briefly summarize the well-known theoretical background of the GW emissions during the inspiralling phase of circular compact binary systems, in order to set the stage for the analysis of the restricted three-body problem, presented in Sec.~\ref{sec:CR3BP}. Then, in Sec.~\ref{sec:validity} we numerically study the validity of our approximations, in particular how large the mass of the third body can become. Finally, we discuss our findings in Sec.~\ref{sec:discussion} and we conclude in Sec.~\ref{sec:conclusions}.

\section{Theoretical background \label{sec:binarysection}}

In this section, we present a brief overview of the GW emission during the inspiralling phase of circular compact binary systems. We recall the principal expressions and formulas involved for better comparison with the results we obtain in the CR3BP case.

The objects forming the binary system are treated as point-like masses, $m_1$,$m_2$, an approximation that is good enough when the separation between the objects is ``large" compared to the Schwarzschild radii involved. Along the same line of reasoning, the background metric can be assumed to be flat, i.e. given by $\eta_{\mu\nu}$, that way all the linearized gravity formalism can be applied. In such a context assuming the orbits of the bodies involved follow the Kepler laws is sufficiently accurate and corrections arising from GR effects are subdominant.\footnote{PN corrections to the orbits are parametrized by the quantity $R_s/r$, which is assumed to be small during the inspiralling phase. For a more precise treatment of the subject, see Ref.~\cite[\S36.7 \S36.8]{Misner}.}Finally, there is the approximation of circularity, which is usually justified by working within the quasi-circular/semi Keplerian regime, where the following applies:
\begin{equation}
    \dot{\omega}_s\ll \omega_s^2 \iff \dot{R}/(\omega_s\,R)\ll 1,
\end{equation}
that is, the tangential velocity is much greater than the rate of contraction of the radius, which means that the approximation of circular orbits with slowly varying radius applies.

It is important to remark that the previous assumptions and following discussions will definitely not apply during the end of the inspiralling phase or merger phase of the binary, where the analytical treatment of the problem becomes intractable and Numerical Relativity is the way to proceed.

The dynamics of the gravitationally bound two body problem is a typical and well studied problem in classical mechanics. Newton's laws provide the differential equations governing the motion of the system:
\begin{equation}\label{Newton_equations}
    m_1\,\bm{\Ddot{r}_1}=-\frac{G\,m_1\,m_2(\bm{r_1-r_2})}{||r_1-r_2||^3},\hspace{1cm}m_2\,\bm{\Ddot{r}_2}=-\frac{G\,m_1\,m_2\,(\bm{r_2-r_1})}{||r_2-r_1||^3}.
\end{equation}
Defining the binary separation $\bm{r}=\bm{r_1}-\bm{r_2}$ and center of mass (CM) vector as $\bm{R}=(m_1\bm{r_1}+m_2\bm{r_2})/(m_1+m_2)$, Eq.~\eqref{Newton_equations} can be rewritten as
\begin{equation}\label{reduction}
    \mu\bm{\Ddot{r}}= -\frac{G\,M\,\mu\,\bm{r}}{||r||^3},\hspace{1cm} \bm{\Ddot{R}}=0,
\end{equation}
where $M=m_1+m_2$ is the total mass of the system and $\mu=m_1\,m_2/M$ is the so called reduced mass. What Eq.~\eqref{reduction} shows is that this two body problem can be reduced to the study of an effective one body problem of mass $\mu$ under the gravitational influence of a mass $M$, by going to the CM frame ($\bm{\Ddot{R}}=0$ guaranteed that way).

In the case of a circular orbit with binary separation $R_b$ in the XY plane, the equations for the orbit of the effective one body are given by
\begin{equation}\label{binary_trajectory}
\begin{cases}
    x(t)=R_b\cos(\omega_s t+\pi/2), \\ y(t)=R_b\sin(\omega_s t+\pi/2), \\z(t)=0.
\end{cases}
\end{equation}
Giving an initial phase of $\pi/2$ corresponds to shifting the origin of time, which is arbitrary, and is set in hindsight to simplify future computations regarding the generation of GWs. Moreover, the radius $R_b$ and angular frequency $\omega_s$ of the orbit are related to each other via Kepler's Third Law:
\begin{equation}\label{kepler}
    \omega_s^2=\frac{G\,M}{R_b^3}\hspace{1mm}.
\end{equation}

Thus, expressions such as the total energy of the system can  equivalently be written either in terms of $R_b$ or $\omega_s$: 
\begin{equation}\label{binary_energy}
    E=E_\mathrm{kin}+E_\mathrm{pot}=-\frac{G\,m_1\,m_2}{2R_b}=-\frac{\mu}{2}(G\,M\,\omega_s)^{2/3}.
\end{equation}
When the different relevant quantities are expressed in terms of $\omega_s$ the factors with mass units, namely $\mu$ and $M$, always happen to appear in the form of powers of the so called chirp-mass:
\begin{equation}
    M_c = \mu^{3/5}\,M^{2/5}=\frac{(m_1\,m_2)^{3/5}}{(m_1+m_2)^{1/5}}\hspace{1mm}.
\end{equation}

The chirp mass is a key quantity to characterize compact binary systems and one of the central parameters to study the distribution of the population of binaries. One of the advantages of working with the chirp mass is that it is relatively straight forward to determine it experimentally. However, to disentangle the individual values of $m_1,m_2$ from the experimental value of $M_c$ complementary measurements or PN corrections are needed.

Taking the orbit given in Eq.~\eqref{binary_trajectory} the quadrupole tensor of the system can be computed, which results in
\begin{equation}\label{derivation_quadrupole_binary}
    M_{ij}(t)=\frac{\mu\, R_b^2}{2}\begin{pmatrix}
        1-\cos(2w_s t) & -\sin(2w_s t)  \\
        -\sin(2w_s t) & 1+\cos(2w_s t)        
    \end{pmatrix} \hspace{1mm},
\end{equation}
where only $i,j=1,2$ are shown, as the rest of the components vanish. Notice that $M_{\hspace{1mm}k}^k=\mu R_b^2$, so upon double differentiation with respect to time the reduced quadrupole tensor $\Ddot{Q}_{ij}$ matches $\Ddot{M}_{ij}$ in the quasi-circular motion approximation. Then, according to the quadrupole formula,
\begin{equation}
   h_{ij}^{TT}(t)=\frac{2G}{D\,c^4}\frac{\mu\, R_b^2}{2}(2\omega_s)^2\begin{pmatrix}
        \cos(2w_s t_r) & \sin(2w_s t_r)  \\
        \sin(2w_s t_r) & -\cos(2w_s t_r)        
    \end{pmatrix}\hspace{1mm},
\end{equation}
which based on the usual definitions of the polarizations $h_+$ and $h_\times$ suggests the identification
\begin{equation}
    h_+(t) = \frac{4G\,\omega_s^2\,\mu\, R_b^2}{D\,c^4}\cos(2\omega_s t_r),\hspace{1cm} h_\times(t) = \frac{4G\,\omega_s^2\,\mu\, R_b^2}{D\,c^4}\sin(2\omega_s t_r),
\end{equation}
where $D$ is the distance between the binary system and the observer.

One of the first important observations is that the frequency of the GWs turns out to be twice the frequency of the source: $\omega_\mathrm{gw}=2\omega_s$. Another important observation which has been ignored so far is the effect of the relative orientation between the source-plane and the observer. So far it has been assumed that the GWs traveled in the $z$ direction, while the motion of the binary was restricted to the XY plane. In a more general setting, where the normal vector to the orbit's plane makes an angle $\theta$ with the line of sight to the observer, it can be shown that the polarization amplitudes pick up the following geometric factors:\footnote{See, for instance, Ref.~\cite[p. 110]{maggiore1}}\begin{eqnarray}\label{polarizations}
h_+(t) &= \frac{4G\,\omega_s^2\,\mu\, R_b^2}{D\,c^4}\,\cos(2\omega_s t_r)\left(\frac{1+\cos^2(\theta)}{2}\right),\nonumber\\
    h_\times(t) &= \frac{4G\,\omega_s^2\,\mu\, R_b^2}{D\,c^4}\,\sin(2\omega_s t_r)\cos(\theta).
\end{eqnarray}

The dependence with respect to the azimuthal angle $\phi$ can be removed by appropriate phase redefinitions and symmetry considerations. An important remark is that both polarizations share the same amplitude $A(t_r)$, which in terms of $f_\mathrm{gw}$ is given by
\begin{equation}\label{commonamplitude}
    A(t_r)=\frac{4}{D}\left(\frac{G\,M_c}{c^2}\right)^{5/3}\left(\frac{\pi f_\mathrm{gw}(t_r)}{c}\right)^{2/3}\hspace{1mm}.
\end{equation}
Regarding the energetic side, the total power radiated by the binary through GWs can either be obtained using $P_\mathrm{quad}=\frac{G}{5c^5}\hspace{1mm}\langle \dddot{Q}_{ij}\,\dddot{Q}_{ij}\rangle\hspace{1mm}$,\footnote{The angular brackets in this scenario effectively correspond to averaging over a period of the orbit.} or integrating $\frac{dP}{d\Omega}=\frac{D^2\,c^3}{16\pi \,G}\hspace{1mm} \langle \dot{h}_+^2+\dot{h}_\times^2\rangle\hspace{1mm}$, both yielding:
\begin{equation}\label{binary_power}
    P_\mathrm{quad}=\frac{32G\,\mu^2\,R_b^4\,\omega_s^6}{5c^5}=\frac{2^{5/3}}{5G\,c^5}(G\,M_c\,\omega_\mathrm{gw})^{10/3}\hspace{1mm}.  
    \end{equation}
Then, energy conservation demands that $P=-\dot{E}$, which using Eq.~\eqref{binary_power} and differentiating Eq.~\eqref{binary_energy} with respect to time translates into the following differential equation:
\begin{equation}\label{freq_variation}
    \dot{f}_\mathrm{gw}=\frac{96}{5}\hspace{1mm}\pi^{8/3}\left(\frac{G\,M_c}{c^3}\right)^{5/3}f_\mathrm{gw}^{11/3},
\end{equation}
The solution to this equation formally diverges for a finite value of the time $t$, which we denote by $t_\mathrm{coal}$, and so it is common to give the solution in terms of the \textit{time to coalescence}, defined as $\tau=t_\mathrm{coal}-t$:
\begin{equation}\label{freq_solution}
    f_\mathrm{gw}(\tau)=\frac{1}{\pi} \left(\frac{G\,M_c}{c^3}\right)^{-5/8}\left(\frac{5}{256}\frac{1}{\tau}\right)^{3/8}.
\end{equation}

The only free parameter in Eq.~\eqref{freq_solution} is the chirp mass $M_c$, so this equation can be used to infer the value of $M_c$ from experimental data (see Ref.~\cite[p.3]{firstGW}). The difference in orbital energy at two different radii $R_1,R_2$ with $R_1>R_2$ is given, according to Eq.~\eqref{binary_energy}, by
\begin{equation}\label{energy_diff}
    \Delta E = \frac{G\,M\,\mu}{2}\left(\frac{1}{R_2}-\frac{1}{R_1}\right)=\frac{\pi^{2/3}}{2G}(G\,M_c)^{5/3}(f_2^{2/3}-f_1^{2/3})\hspace{1mm},
\end{equation}
where $f_1,f_2$ with $f_1<f_2$ are the GW frequencies corresponding to $R_1,R_2$ through Kepler's Third Law. This result follows from purely classical arguments, but should match the energy lost through emission of GWs. To compute such a quantity the differential energy spectrum $dE/df$ might be used:
\begin{equation}\label{diff_energy_spectrum}
    \frac{dE}{df}=\frac{\pi c^3}{2G}f^2\,D^2\int d\Omega\left(|\Tilde{h}_+(f)|^2+|\Tilde{h}_\times(f)|^2\right)\hspace{1mm},   
\end{equation}
for which the Fourier transforms of the polarization amplitudes need to be computed. To do so, although the quasi-circular motion approximation basically assumes a constant angular frequency at any given time, the phase term $\omega_\mathrm{gw}t$ needs to be replaced by the more accurate\footnote{For a more careful analysis, see Ref.~\cite[p.172]{maggiore1}.} phase term 
\begin{equation}
    \Phi(t)=\int_{t_0}^t\omega_\mathrm{gw}(u)du\hspace{1mm}.
\end{equation}

Plugging the solution Eq.~\eqref{freq_solution} into this phase term and expressing it in terms of the time to coalescence $\tau$:

\begin{equation}\label{bigphase}
    \Phi(\tau)=\int^0_{\tau}\omega_\mathrm{gw}(u)du=-2\left(\frac{5G\,M_c}{c^3}\right)^{-5/8}\tau^{5/8}+\Phi_0\hspace{1mm},
\end{equation}
\phantom{a}\\
where $\Phi_0$ is the phase at coalescence.

Computing the Fourier transforms of the polarization amplitudes is not straight-forward, given that the functions $h_+(t),~h_\times (t)$ are only defined for $-\infty<t<t_\mathrm{coal}$ and not over the entire real line. The usual procedure is to make use of the stationary phase approximation, a standard tool in the study of asymptotic behavior of integrals (similar to the method of steepest descent). The main idea behind this method is to disregard contributions to the integral coming from complex exponentials with rapidly varying phases and similar amplitudes, due to the fact that they will add incoherently and cancel each other out~\cite[p.76]{asymptotic}. More technically, the integrand is expanded to second order around its stationary points, i.e. points where the derivative (1D case) of the integrand vanishes. Explicitly:
\begin{equation}
\begin{aligned}
    \Tilde{h}_+(f)&=\int dt A(t_r)\cos(\Phi(t_r))e^{i2\pi ft} \\
    &=\frac{1}{2}e^{i2\pi f\frac{D}{c}}\int dt A(t_r)\left[e^{i\Phi(t_r)}+e^{-i\Phi(t_r)}\right]e^{i2\pi ft_r} \\
    &=\frac{1}{2}e^{i2\pi f\frac{D}{c}}\int dt A(t)\left[e^{i(\Phi(t)+2\pi f t)}+e^{i(2\pi f t -\Phi(t))}\right].    
    \end{aligned}
\end{equation}

Noticing that $\dot{\Phi}(t)=\omega_\mathrm{gw}>0$, the first exponential does not have any stationary points and will always oscillate fast, while the the second exponential, assuming that $\log A(t)$ varies slowly compared to $\Phi(t)$, has a stationary point $t_s(f)$ determined by the condition \footnote{This comes as no surprise: it basically comes to say that the dominant contribution for the frequency $f$ comes from values of time close to the time for which $\omega_\mathrm{gw}=2\pi f$.}

\begin{equation}
   \frac{d}{dt}\left( 2\pi f t- \Phi(t) \right)\Big\rvert_{t=t_s}=0 \implies 2\pi f = \dot{\Phi}(t_s)\hspace{1mm}.
\end{equation}
Then:
\begin{equation}
  \Tilde{h}_+(f) \approx \frac{1}{2}e^{i2\pi f\frac{D}{c}}\int dt A(t) e^{i(2\pi f t -\Phi(t_r))}  .
\end{equation}
Expanding the exponential to second order around $(t-t_s)$ and using the stationary point condition:
\begin{equation}\label{stationary_result}
\begin{aligned}
 \Tilde{h}_+(f)&\approx\frac{1}{2}e^{i2\pi fD/c}\,A(t_s)\,e^{i[2\pi ft_s-\Psi(t_s)]}\,\sqrt{\frac{2\pi}{\Ddot{\Phi}(t_s)}}\,\int_{-\infty}^{\infty}e^{-iu^2}\hspace{1mm}du\\
 &=\frac{1}{2}e^{i\Psi_+}\,A(t_s)\,\sqrt{\frac{2\pi}{\Ddot{\Phi}(t_s)}}\hspace{1mm},
\end{aligned}    
\end{equation}
where the last integral is a Fresnel integral evaluating to $\sqrt{\pi}e^{-i\pi/4}$ and all the complex exponential arguments have been grouped into
\begin{equation}
    \Psi_+(f)=2\pi f\left(t_s+\frac{D}{c}\right)-\Phi_0-\frac{\pi}{4}+\frac{3}{4}\left(\frac{G\,M_c}{c^3}\hspace{1mm}8\pi f\right)^{-5/3}.
\end{equation}
Finally, using Eq.~\eqref{commonamplitude} and differentiating Eq.~\eqref{bigphase} twice, the following expression is obtained:
\begin{equation}\label{binary_fourier}
    \Tilde{h}_+(f)\approx \frac{1}{\pi^{2/3}}\left(\frac{5}{24}\right)^{1/2}e^{i\Psi_+}\hspace{1mm}\frac{c}{D}\left(\frac{G\,M_c}{c^3}\right)^{5/6}\frac{1}{f^{7/6}}\left(\frac{1+\cos^2\theta}{2}\right)\hspace{1mm}.
\end{equation}
An almost identical derivation yields the result for the $\times$ polarization:
\begin{equation}
    \Tilde{h}_\times(f)\approx \frac{1}{\pi^{2/3}}\left(\frac{5}{24}\right)^{1/2}e^{i\Psi_\times}\hspace{1mm}\frac{c}{D}\left(\frac{G\,M_c}{c^3}\right)^{5/6}\frac{1}{f^{7/6}}\cos\theta\hspace{1mm},
\end{equation}
where $\Psi_\times=\Psi_++\frac{\pi}{2}$, due to an extra $i$ factor. Substituting these two results into Eq.~\eqref{diff_energy_spectrum} and integrating from $f=f_1$ to $f=f_2$, one gets 
\begin{equation}\label{radiated_inspiral}
    \Delta E_\mathrm{insp}=\frac{\pi^{2/3}}{2G}(GM_c)^{5/3}(f_2^{2/3}-f_1^{2/3})\hspace{1mm},
\end{equation}
which perfectly matches the prediction in Eq.~\eqref{energy_diff}.

\section{The Circular Restricted Three Body Problem \label{sec:CR3BP}}

The CR3BP assumes that the masses of the two larger bodies are much greater than that of the third body, and therefore that, at the classical level, the orbit of the binary is unaffected by its presence. An extensive description of the classical CR3BP can be found in Ref.~\cite[Ch.8]{celestial}. In this problem, there are five equilibrium points, known as Lagrange points, which play a crucial role in the dynamics of the system. A widely studied example of the CR3BP is the Sun-Jupiter system, which acts as the parent binary system for the numerous Trojan satellites congregating around the $L_4$ and $L_5$ Lagrange points. When GR effects are taken into account the orbital parameters governing the dynamics of the problem are modified~\cite{PNcorrections,3bodyastro}, but they only become relevant towards the end of the inspiralling phase. 

The purpose of this section is to study the GW emission spectrum of such a system during the inspiralling phase when extremely compact objects such as BHs and NSs are involved. In what follows, we assume semi-keplerian orbits for the bodies involved. After computing the polarizations of the emitted GWs, we account for the change in orbital parameters (mainly the decrease in orbital radius, and therefore, the increase in orbital frequency) due to the orbital energy loss associated to the radiated GW energy. We then compare the obtained expressions with the analogous results for binaries and discuss the observational implications that resonating third bodies could have. 

\begin{figure}[!t]
    \centering
    \includegraphics[scale=0.6]{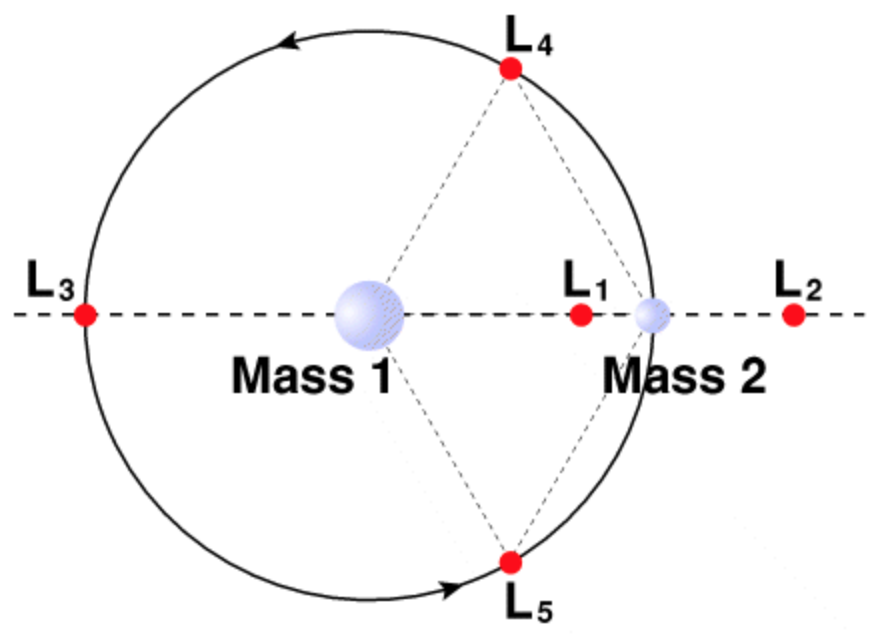}
    \caption{Position of the Lagrange points $L_1$ to $L_5$. Source: NASA.}
    \label{fig:lagrangepoints}
\end{figure}

In any binary system, there are certain special points where the gravitational forces from the two bodies balance in such a way that a third, smaller body can maintain a stable position relative to the binary system: the Lagrange points. For any setup, there are five such points~\cite[p.48]{celestial}, denoted by $L_1$ to $L_5$, which are illustrated in Fig.~\ref{fig:lagrangepoints}.

Out of the 5 Lagrange points only two of them can in principle admit stable orbits around them: Lagrange points $L_4$ and $L_5$. Orbits sitting at the other three Lagrange points are not stable and require external energy (provided by thrusters or engines, for instance) to orbit in 1:1 resonance with the parent binary. 

The stability of the orbits at $L_4$ and $L_5$ depends on the mass ratio
\begin{equation}
    \eta=\frac{m_2}{m_1+m_2}\leq0.5\hspace{1mm}.
\end{equation}
Specifically, it can be shown that for orbits to be stable $\eta\leq \eta_\mathrm{crit}$ is needed~\cite[\S 8.6]{celestial}, where
\begin{equation}\label{etacrit}
    \eta_\mathrm{crit}= 1/2-\sqrt{69}/18\approx 0.03852\hspace{1mm}.
\end{equation}
In such cases, the motion around the Lagrange points can be described by small oscillations with frequencies
\begin{equation}\label{epicycles}
    \omega_{1,2}=\omega_s\sqrt{\frac{1\pm\sqrt{1-27\eta\,(1-\eta)}}{2}}\hspace{1mm},
\end{equation}
where $\omega_s$ is the angular frequency of the parent binary. It must be noted that these frequencies are a result of a perturbative expansion and there are no general exact analytic solutions for the perturbed quasi-periodic orbits~\cite[p.174]{celestial}.

\begin{figure}[!t]
    \centering
    \includegraphics[scale=0.32]{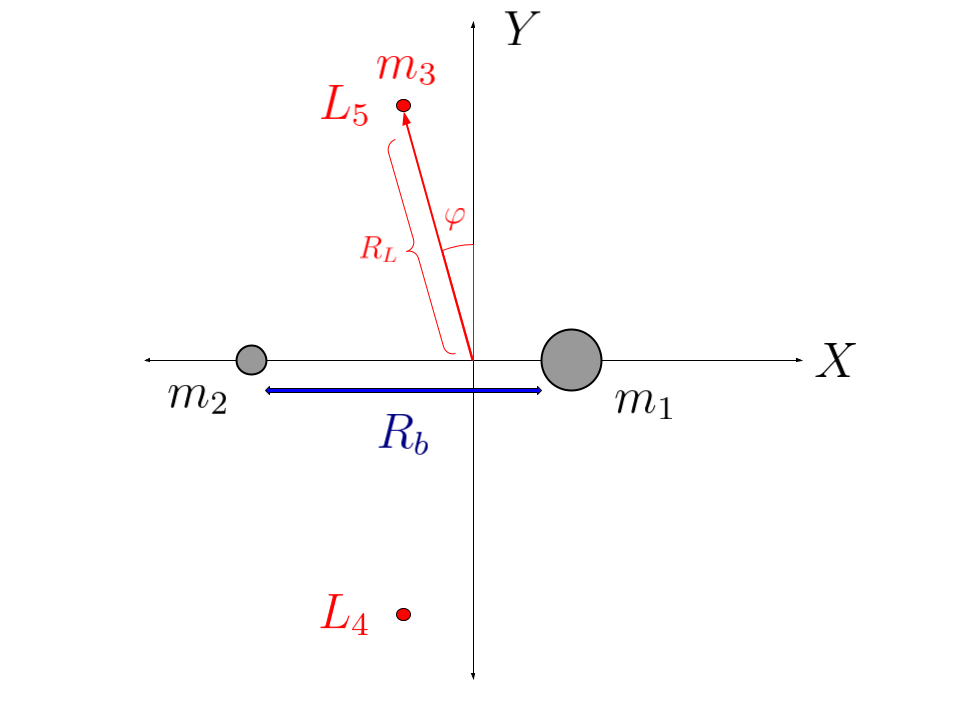}
    \caption{Geometrical setup of the CR3BP.}
    \label{fig:sketch}
\end{figure}

\begin{figure}[!t]
    \centering
    \includegraphics[scale=0.4]{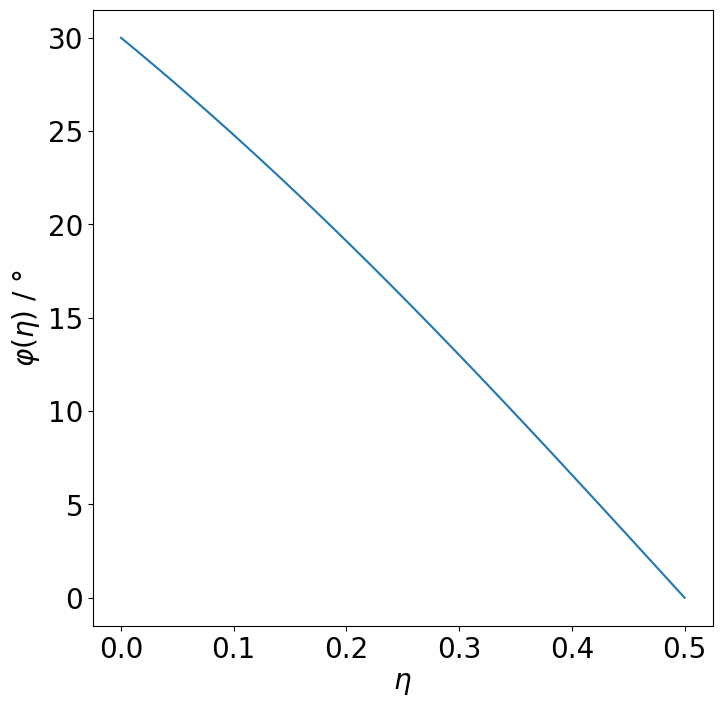}
    \caption{Value of the angle $\varphi$ subtended by the Lagrange point as shown in Fig.~\ref{fig:sketch} in terms of the mass asymmetry $\eta$.}
    \label{fig:phi function}
\end{figure}

The following discussion always assumes that the condition $\eta<\eta_\mathrm{crit}$ is fulfilled, indicating significant mass asymmetry of the parent binary system. In Fig.~\ref{fig:sketch}, we diagrammatically show the geometrical setup of the problem that is studied in this section, represented in the co-rotating binary CM frame. The position of the relevant Lagrange points is given, in this coordinate system, by:
\begin{equation}
    X_L=\left(\eta-\frac{1}{2}\right)R_b,\hspace{15mm}Y_L=\pm\frac{\sqrt{3}}{2}R_b,
\end{equation}
and the distance from the origin to either of the Lagrange points is
\begin{equation}
    R_L=\sqrt{X_L^2+Y_L^2}=\sqrt{1-\eta+\eta^2}\hspace{1mm}R_b\hspace{1mm}.
\end{equation}
The angle $\varphi$ in Fig.~\ref{fig:sketch} is a function of the mass ratio $\eta$: 
\begin{equation}
    \varphi(\eta) = \arctan\left(\frac{|X_L|}{Y_L}\right)=\arctan\left(\frac{1-2\eta}{\sqrt{3}}\right)\hspace{1mm},
\end{equation}
while the values taken by $\varphi(\eta)$ for $\eta\in [0,0.5]$ are shown in Fig.~\ref{fig:phi function}.
The maximum value taken by $\varphi$ is when $\eta\to 0$, the exact value being $\varphi(0)=\pi/6=30^\circ$. The positions of $m_1,m_2$ in the co-rotating binary CM reference frame are $R_b(\eta ,0)$ and $R_b(\eta-1,0)$, respectively, which implies that indeed $m_1,m_2$ and $m_3$ form an equilateral triangle as long as $m_3$ sits exactly at the Lagrange point. 
 
As far as the dynamics of the system are concerned, the main classical assumption of the CR3BP is that the binary system evolves unaffected by the presence of the third body. In practical terms, this means that Eqs.~\eqref{binary_trajectory} and ~\eqref{kepler} are good Keplerian approximations to the orbits of the binary. Moreover, the geometrical configuration shown in Fig.~\ref{fig:sketch} is assumed to hold at every time, even as the value of $R_b$ changes. Therefore, the equations for the third body's orbit in the binary's CM frame are 
\begin{equation}\label{3trajectory}
\begin{cases}
    x(t)=R_L\cos(\omega_s t+\pi+\varphi),\\ y(t)=R_L\sin(\omega_s t+\pi+\varphi), \\z(t)=0.
\end{cases}
\end{equation}

The extra $\pi$ in the phase is introduced so that the origin of time matches the convention adopted for Eq.~\eqref{binary_trajectory}. Also, given that in the approximation of considering the bodies as point masses the energy-momentum tensor is additive, the total quadrupole moment of the system will just be the sum of the quadrupole moments coming from the binary and the 3rd body.
\begin{equation}
    Q_\mathrm{total}=Q_\mathrm{binary}+Q_3\hspace{1mm}.
\end{equation}
Here, some caution is needed if the quadrupole approximation is to be used. As highlighted in~\cite{Bonetti:2017hnb}, the system must be within the so-called Near Coordinate Zone (NCZ), a region from the CM with a radius comparable to wavelength of the GWs at play, for the current formalism to remain valid. However, given the geometry and assumptions of the CR3BP, where the distance from the CM to the third body is comparable to the binary separation $R_b$, the NCZ condition can easily be satisfied in our system by restricting it to the usual inspiral phase.

The quadrupole moment corresponding to the orbital trajectory for the third body given in Eq.~\eqref{3trajectory} is derived in an exactly analogous manner to Eq.~\eqref{derivation_quadrupole_binary}, which according to the quadrupole formula then yields
\begin{eqnarray}
    \hspace{-2cm}h_+(t) =& \frac{4G\omega_s^2\mu R_b^2}{D\,c^4}\left[ \cos(2\omega_s t_r)-(1-\eta+\eta^2)\left( \frac{m_3}{\mu}\right) \cos(2\omega_s t_r\pm2\varphi)\right]\left(\frac{1+\cos^2\theta}{2}\right),\nonumber\\
    \hspace{-2cm}h_\times(t) =&\frac{4G\omega_s^2\mu R_b^2}{D\,c^4}\left[ \sin(2\omega_s t_r)-(1-\eta+\eta^2)\left( \frac{m_3}{\mu}\right) \sin(2\omega_s t_r\pm2\varphi)\right]\cos\theta.
    \label{polarizations3}
\end{eqnarray}

From this, we observe that the contribution from the third body corresponds to the second term in square brackets. The term $(1-\eta+\eta^2)$ is expected to be close to one, given the stability constraint imposed by Eq.~\eqref{etacrit}. The term $m_3/\mu$ is also expected to be small, in order to fall within the assumptions of the CR3BP regime.

\begin{figure}[!t]
    \centering
    \includegraphics[scale=0.07]{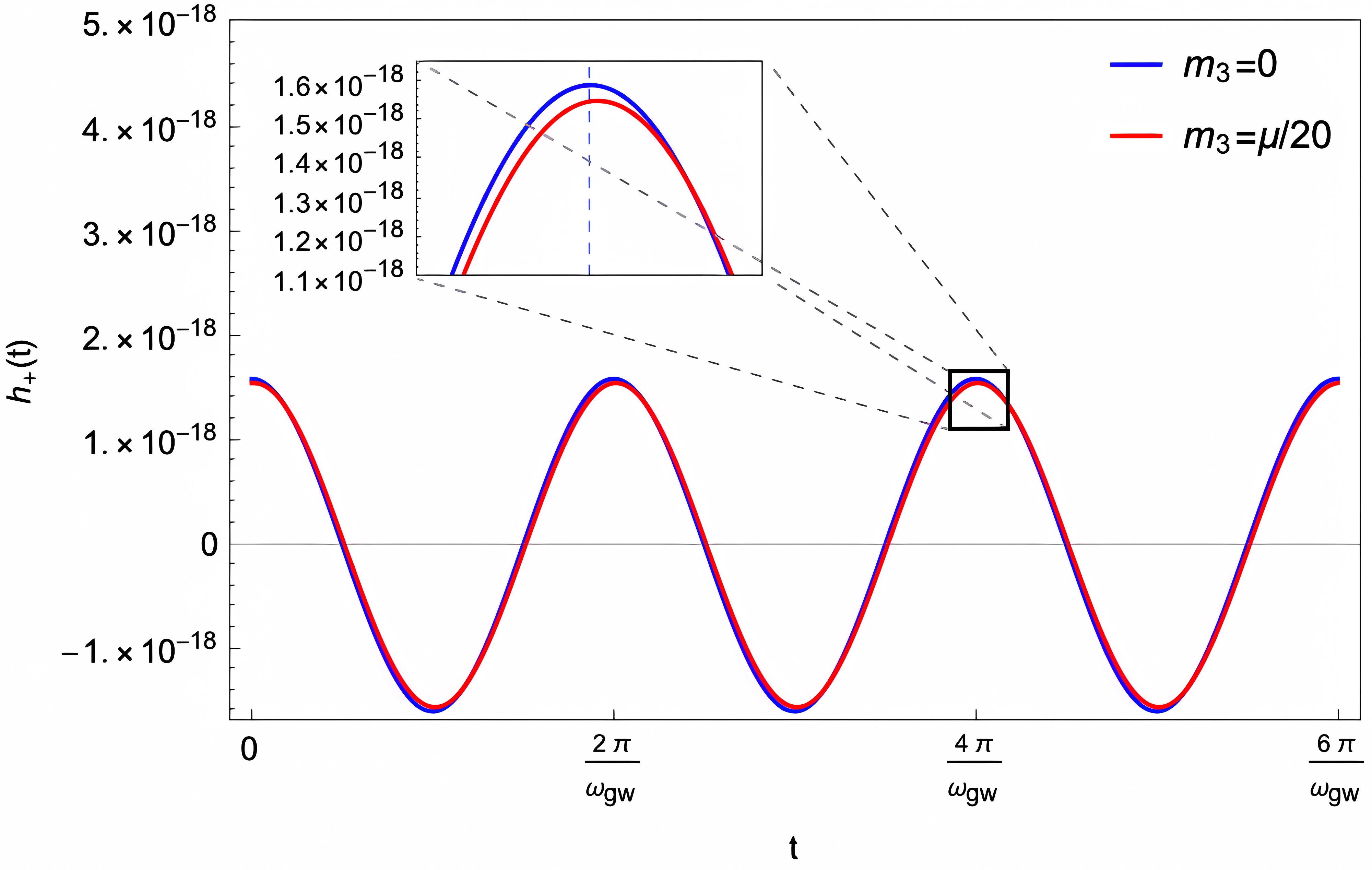}
    \caption{Comparison of the $h_+$ polarization amplitude between a binary system (blue) and a CR3BP system obtained by adding a small third object (red). We assume a binary located $100$Mpc from the observer, with masses $m_1=2\times 10^6M_{\odot}$, $m_2=5\times 10^4M_{\odot}$ and binary separation $R_b=60R_{s,m_1}$, resulting in $f_{\rm gw}\approx 7\times 10^{-5}$Hz.}
    \label{fig:polarizations}
\end{figure}

In Fig.~\ref{fig:polarizations}, we show the GW waveform with and without the small third body to illustrate the effect (the plot for $h_\times$ is essentially the same). The effect can be interpreted as a reduction in the amplitude and an overall phase shift. In fact, since the sum of two phasors with the same frequency amounts to a third phasor with the same frequency, making use of trigonometric identities, Eq.~\eqref{polarizations3} can be rewritten as
\begin{align}
   h_+(t) =& A_+\cos(2\omega_s t_r+\varphi_0)\left(\frac{1+\cos^2\theta}{2}\right),\nonumber\\
   h_\times(t) =&A_\times\sin(2\omega_s t_r+\varphi_0)\cos\theta,
    \label{phasor}
\end{align}
where 
\begin{align}
  A_+&=A_\times= \frac{4G\omega_s^2\mu R_b^2}{D\,c^4}\left[1-(1-\eta+\eta^2)\left( \frac{m_3}{\mu}\right) \right],\\\varphi_0&=\mp\tan^{-1}\left[\frac{(1-\eta+\eta^2)\left( \frac{m_3}{\mu}\right)\sin(2\varphi) }{1-(1-\eta+\eta^2)\left( \frac{m_3}{\mu}\right)\cos(2\varphi) }\right] ,
\end{align}

which facilitates the interpretation of Fig.~\ref{fig:polarizations}.

A more detailed analysis regarding the feasible magnitude of the mass ratio is provided in Section \ref{sec:validity}, but an important realization to make is that given the high mass asymmetry imposed on the binary one has $\mu\sim m_2$, and so a value of $m_3/\mu=0.1$ would already mean that $m_3$ is very small compared to the total mass $M=m_1+m_2$.
    
Back to Eq.~\eqref{polarizations3}, the phase of the third body's polarization is shifted with respect to the binary's by $2\varphi$. This will have some important consequences. The $\pm$ sign arises from simultaneously considering the possibilities of the third body sitting in either $L_4$ or $L_5$. Also, the polarizations have the same angular dependence with respect to observer as in the binary case.

To study the potential effect of the third body in the dynamical evolution of the system due to its GW effects, we begin by computing the power $P_\mathrm{3B}$ emitted in the form of GWs in the CR3BP: 
\begin{equation}\label{3power}
    P_\mathrm{3B}=P_\mathrm{2B}\left(\hspace{1mm}1-2\cos(2\varphi)(1-\eta+\eta^2)\left(\frac{m_3}{\mu}\right)+(1-\eta+\eta^2)^2\left(\frac{m_3}{\mu}\right)^2\hspace{1mm}\right)\hspace{1mm},
\end{equation}
where we have used that
\begin{equation}
    \langle\sin(2\omega_st)\sin(2\omega_st+2\varphi)\rangle=\langle\cos(2\omega_st)\cos(2\omega_st+2\varphi)\rangle=\frac{\cos(2\varphi)}{2}\,,
\end{equation}
and $P_\mathrm{2B}$ is the binary power expression given in Eq.~\eqref{binary_power}. The interesting comment regarding Eq.~\eqref{3power} is that, for realistic values of $\eta$ and $m_3/\mu$, one has that $P_\mathrm{3B}<P_\mathrm{2B}$, since the interference term dominates over the third body's term. Thus, one of the first interesting results regarding GWs in the CR3BP is that energy is generally lost at a slower rate.

The energy lost through the emission of GWs must come from the available orbital energy, which can be written in terms of just the binary separation $R_b$:
\begin{equation}
    E_\mathrm{3B}(R_b)=-\frac{G\,m_1\,m_2}{2R_b}-\frac{G\,m_1\,m_3}{R_b}-\frac{G\,m_2\,m_3}{R_b}+E_\mathrm{kin}^{(3)}\hspace{1mm}.
\end{equation}
\phantom{a}\\
For the kinetic energy of third body its velocity is needed. Since, under the assumptions adopted, it co-rotates with the binary:
\begin{equation}
    v=\omega_s R_L=\sqrt{\frac{G\,M}{R_b^3}}\,R_L=\sqrt{\frac{G\,M(1-\eta+\eta^2)}{R_b}}\hspace{1mm},
\end{equation}
which implies that
\begin{equation}
    E_\mathrm{kin}^{(3)}=\frac{m_3}{2}\frac{G\,M}{R_b}(1-\eta+\eta^2)\hspace{1mm}.
\end{equation}
After some simple algebra, the following expression for the total orbital energy of the system is obtained
\begin{equation}\label{3energy}
    E_\mathrm{3B}=E_\mathrm{2B}-\frac{G\,m_3\,M}{2R_b}(1+\eta-\eta^2)=E_\mathrm{2B}\left(1+(1+\eta-\eta^2)\left(\frac{m_3}{\mu}\right)\right).
\end{equation}
The term inside the parenthesis is always greater than one, which basically comes to remark the rather natural fact that the total energy of the system is higher (more negative) than that of the binary on its own. Writing Eq.~\eqref{3power} and Eq.~\eqref{3energy} in terms of $P_\mathrm{2B}$ and $E_\mathrm{2B}$, respectively, has the advantage of simplifying upcoming calculations, since many results concerning the only-binary case shown in Sec.~\ref{sec:binarysection} will apply.

\begin{figure}[!t]
    \centering
    \captionsetup{width=.9\linewidth}
    \includegraphics[width=.9\linewidth]{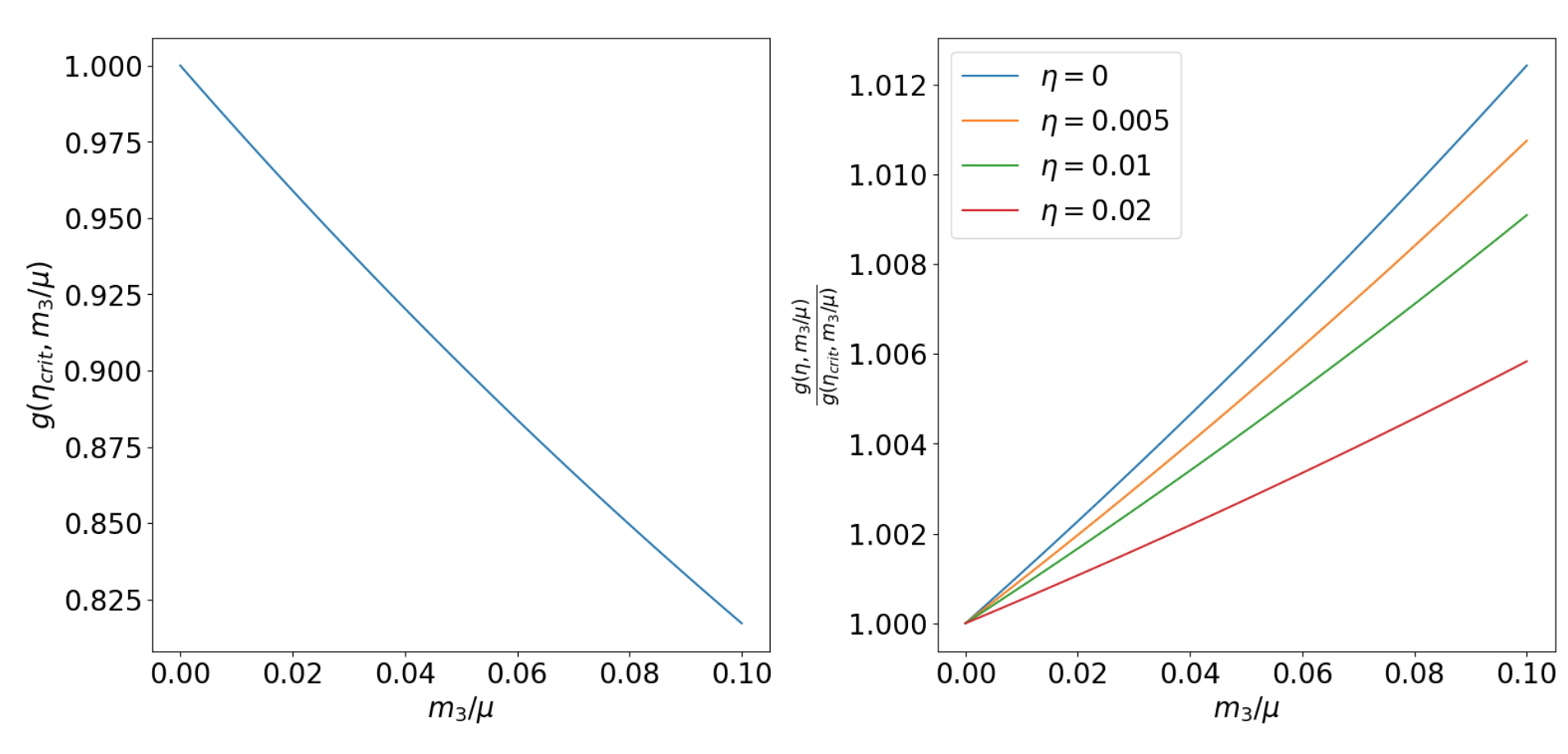}
    \caption{Left: The $g(\eta_\mathrm{crit},m_3/\mu)$ function in the range $0<m_3/\mu<0.1$. Right: Relative variation of the $g$ function in terms of $\eta$.}
    \label{fig:g_function}
\end{figure}

Imposing the energy conservation condition $P_\mathrm{3B}=-\dot{E}_\mathrm{3B}$ gives the differential equation governing the frequency evolution: 
\begin{equation}\label{new_freq}
    \dot{f}_\mathrm{gw}=\frac{96}{5}\hspace{1mm}\pi^{8/3}\left(\frac{GM_c}{c^3}\right)^{5/3}f_\mathrm{gw}^{11/3} g(\eta,m_3/\mu),
\end{equation}
where $g(\eta,m_3/\mu)$ is just
\begin{equation}
    g(\eta,m_3/\mu)=\left[\frac{1-2\cos(2\varphi)(1-\eta+\eta^2)(m_3/\mu)+(1-\eta+\eta^2)^2(m_3/\mu)^2}{1+(1+\eta-\eta^2)(m_3/\mu)}\right].
\end{equation}
In Fig.~\ref{fig:g_function}, we show this $g(\eta,m_3/\mu)$ function  for a few different values of $\eta$. A few comments regarding this function:
The variation with $\eta$ is rather small. From the right panel it can be observed that the maximum relative difference is of order $1\%$. Moreover, it is a decreasing function of $m_3/\mu$, the principal variable that determines it, and in the absence of the third body, i.e. $m_3=0$, the expected result $g(\eta,0)=1$ is recovered. This will serve as a guiding principle to perform sanity checks for the rest of results: setting $g(\eta,m_3/\mu)=1$ must recover the results corresponding to the only-binary scenario.

\begin{figure}[!t]
    \centering
    \includegraphics[scale=0.37]{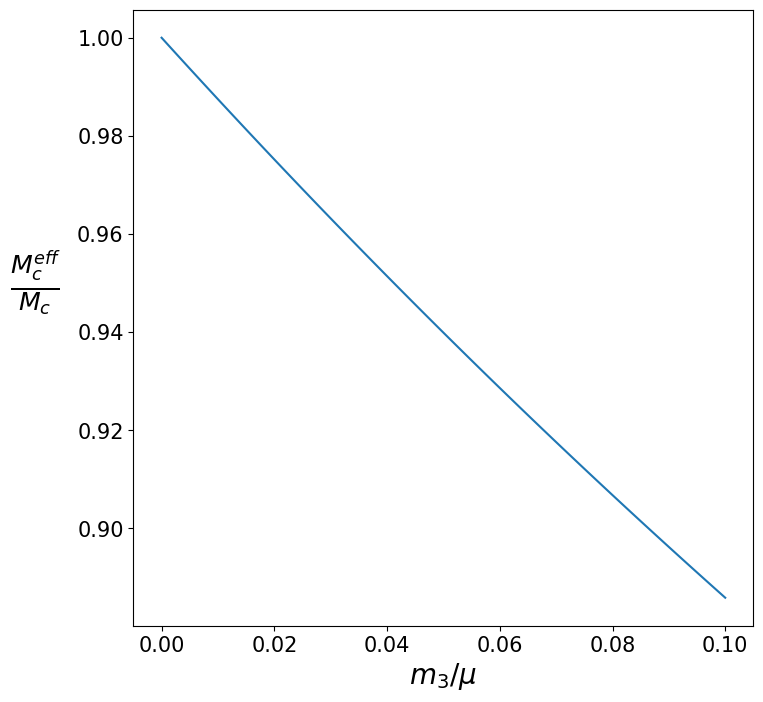}
    \caption{Ratio $M_c^{\text{eff}}/M_c$ in terms of $m_3/\mu$ for $\eta=\eta_\mathrm{crit}$.}
    \label{fig:chirpmass}
\end{figure}

A key observation is that Eq.~\eqref{new_freq} can be interpreted as an effective rescaling of the chirp mass of the binary:
\begin{equation}\label{newchirp}
    M_c^{\text{eff}}=M_c\, g(\eta,m_3/\mu)^{3/5}\hspace{1mm}.
\end{equation}
Since $g(\eta,m_3/\mu)<1$, it follows that $M_c^{\text{eff}}<M_c$, that is, the new effective chirp-mass is smaller that the original one. 
Thus, under the considered approximations, there is a degeneracy in the determination of the chirp mass from the frequency evolution of the GWs: the signal coming from the inspiralling phase of a binary with chirp mass $M_c$ and the signal coming from a CR3BP scenario with a more massive binary but with the same effective chirp mass $M_c^{\text{eff}}$ would be indistinguishable. In Fig.~\ref{fig:chirpmass}, we show how big of a rescaling the third body induces in the chirp mass. 

Figure~\ref{fig:chirpmass} shows that for reasonable values of $m_3/\mu$ the rescaling of the chirp mass can be of the order of $5\%$. Following up, the solution to Eq.~\eqref{new_freq} in terms of the time to coalescence, in analogy to Eq.~\eqref{freq_solution}, is
\begin{equation}\label{new_frequencyy}
\begin{aligned}
    f_\mathrm{gw}(\tau)&=\frac{1}{\pi} \left(\frac{GM_c}{c^3}\right)^{-5/8}\left(\frac{5}{256}\frac{1}{\tau}\right)^{3/8}g(\eta,m_3/\mu)^{-3/8}\\
    &=\frac{1}{\pi} \left(\frac{GM_c^{\text{eff}}}{c^3}\right)^{-5/8}\left(\frac{5}{256}\frac{1}{\tau}\right)^{3/8}.
\end{aligned}
\end{equation}
Equivalently, reversing the previous equation, the time to coalescence will be related to the frequency of the GWs via the expression
\begin{equation}\label{time_to_coalescence}
    \tau = \frac{5}{256}\left(\frac{GM_c^{\text{eff}}}{c^3}\right)^{-5/3}(\pi f_\mathrm{gw})^{-8/3} \hspace{1mm}.
\end{equation}

A physically relevant conclusion is that the presence of the third body slows down the coalescence of the binary: for a given initial frequency $f_0$ (that is to say, an initial binary separation $R_0$), Eq.~\eqref{time_to_coalescence} shows that a smaller chirp mass corresponds to a longer time to coalescence.

We have argued that the effect of having a third small co\-ro\-ta\-ting object is an effective diminishing of the chirp mass associated to the binary system. Strictly following that reasoning, however, would lead to the conclusion that less total energy is emitted during the inspiral phase, at least according to Eq.~\eqref{radiated_inspiral}.

We now demonstrate that this is not the case, as more energy is indeed radiated when the third body is present by the time a certain binary separation $R_0$ is reached. To do so, we compute the frequency differential energy spectrum $dE/df$ for the CR3BP in a similar fashion of Section \ref{sec:binarysection}. For that purpose, we begin by computing the GW polarizations in Fourier space.

The polarizations in the time domain given in Eq.~\eqref{polarizations3} can be rewritten in a more compact way as
\begin{equation*}
    h_+(t)=A(t_r)[\cos(\Phi(t_r))-B\cos(\Phi(t_r)\pm 2\varphi)],
\end{equation*}
\begin{equation}
    h_\times(t)=A(t_r)[\sin(\Phi(t_r))-B\sin(\Phi(t_r)\pm 2\varphi)],
\end{equation}
\phantom{a}\\
where $B=(1-\eta+\eta^2)(m_3/\mu)$ and
\begin{equation}\label{newamplitude}
  A(t_r)=\frac{4}{D}\frac{(GM_c)^{5/3}}{c^4}(\pi f_\mathrm{gw}(t_r))^{11/3} \hspace{1mm},    
\end{equation}
\begin{equation}\label{phaseterm}
    \Phi(\tau) =\int_{\tau}^0 \omega(t')dt'=-2\left(\frac{5GM_c}{c^3}\right)^{-5/8}\tau^{5/8}g(\eta,m_3/\mu)^{-3/8}+\Phi(\tau_0)\hspace{1mm}.
\end{equation}

Then:
\begin{equation}\label{longfourier}
\begin{aligned}
    \Tilde{h}_+(f)&=\int dt A(t_r)\left[\cos(\Phi(t_r))-B\cos(\Phi(t_r)\pm 2\varphi)\right]e^{i2\pi ft} \\
    &=\frac{1}{2}e^{i2\pi f\frac{D}{c}}\int dt A(t_r)\left[e^{i\Phi(t_r)}+e^{-i\Phi(t_r)}-Be^{i\Phi(t_r)}e^{\pm i2\varphi}-Be^{-i\Phi(t_r)}e^{\mp i2\varphi}\right]e^{i2\pi ft_r}\\
    &=\frac{1}{2}e^{i2\pi f\frac{D}{c}}\int dt A(t)\left[e^{i(\Phi(t_r)+2\pi f t)}(1-Be^{\pm i2\varphi})+e^{i(2\pi f t -\Phi(t_r))}(1-Be^{\mp i2\varphi})\right]\\
    &\approx \frac{1}{2}e^{i2\pi f\frac{D}{c}}\int dt A(t) e^{i(2\pi f t -\Phi(t_r))}(1-Be^{\mp i2\varphi})\\
    &=(1-Be^{\mp i2\varphi})\left[\frac{1}{2}e^{i2\pi f\frac{D}{c}}\int dt A(t) e^{i(2\pi f t -\Phi(t_r))}\right].
    \end{aligned}
\end{equation}
From this point on wards the procedure to compute the integral is analogous to that shown in Section \ref{sec:binarysection}, i.e. by making use of the stationary phase approximation. At first glance, one might naively think that this last integral will yield the same result as for the binary case, however, the time evolution of $A(t)$ and $\Phi(t)$ are now given by Eqs.~\eqref{newamplitude} and ~\eqref{phaseterm} respectively, so some extra care needs to be taken.

The term in brackets in the last line of Eq.~\eqref{longfourier} will also be of the form of \eqref{stationary_result}, but taking the second derivative of Eq.~\eqref{phaseterm} and substituting Eq.~\eqref{time_to_coalescence} into it gives an additional factor $g(\eta,m_3/\mu)^{-1/2}$. After performing all the relevant substitutions, the following expression is obtained:
\begin{equation}\label{3fourier}
    \Tilde{h}_+(f)=\frac{(1-Be^{\mp i2\varphi})}{\sqrt{g(\eta,m_3/\mu)}}\left[\frac{1}{\pi^{2/3}}\left(\frac{5}{24}\right)^{1/2}e^{i\Psi_+'}\hspace{1mm}\frac{c}{D}\left(\frac{GM_c}{c^3}\right)^{5/6}\frac{1}{f^{7/6}}\left(\frac{1+\cos^2\theta}{2}\right)\hspace{1mm}\right]\hspace{1mm},
\end{equation}
where
\begin{equation}\label{3psi}
\begin{aligned}
    \Psi_+'(f)&=2\pi f\left(t_s+\frac{D}{c}\right)-\Phi_0-\frac{\pi}{4}+\frac{3}{4}\left(\frac{GM_c}{c^3}\hspace{1mm}8\pi f\right)^{-5/3}g(\eta,m_3/\mu)^{-1}\\
    &=2\pi f\left(t_s+\frac{D}{c}\right)-\Phi_0-\frac{\pi}{4}+\frac{3}{4}\left(\frac{GM_c^{\text{eff}}}{c^3}\hspace{1mm}8\pi f\right)^{-5/3}.
\end{aligned}    
\end{equation}

Performing almost identical calculations it is straightforward to show that the same prefactor appears for $\Tilde{h}_\times(f)$ too and $\Psi_\times'=\Psi_+'+\pi/2$. The term in square brackets in Eq.~\eqref{3fourier} is basically the same as Eq.~\eqref{binary_fourier}, with the small difference of changing $\Psi_+$ for $\Psi_+'$, which in any case according to Eq.~\eqref{diff_energy_spectrum} does not influence the differential energy spectrum. Then,
\begin{equation}
    \frac{dE}{df}=\left|\frac{(1-Be^{\mp i2\varphi})}{\sqrt{g(\eta,m_3/\mu)}}\right|^2\left(\frac{dE}{df}\right)_\mathrm{2B}=\left[1+(1+\eta-\eta^2)\left(\frac{m_3}{\mu}\right)\right]\left(\frac{dE}{df}\right)_\mathrm{2B}.
\end{equation}

Comparing this last expression and the one for the orbital energy of the CR3BP given by Eq.~\eqref{3energy} it follows that the energy radiated away in the form of GWs during the inspiralling phase matches the orbital energy loss:

\begin{equation}\label{acabateya}
    \Delta E_\mathrm{insp}=\left(1+(1+\eta-\eta^2)\left(\frac{m_3}{\mu}\right)\right)\frac{\pi^{2/3}}{2G}(GM_c)^{5/3}(f_2^{2/3}-f_1^{2/3})\hspace{1mm}.
\end{equation}

However, the effect of the third body on the binary cannot be summarized just with the effective chirp mass rescaling: substituting $M_c^{\text{eff}}$ for $M_c$ in Eq.~\eqref{binary_fourier} does not yield Eq.~\eqref{3fourier}. Also, although the difference between $\Psi_+,\Psi_\times$ and $\Psi_+',\Psi_\times'$ does not come into play in the energetic considerations, it can play an important role in the detection of transient GW signals in the data provided by current detectors. Along those lines, Eq.~\eqref{3psi} picks up an extra correction from the $(1-Be^{\mp i2\varphi})$ term if we wish to enclose all the complex part of $\Tilde{h}_+(f)$ within it:
\begin{equation}
        \begin{aligned}
    \Psi_+'(f)&=2\pi f\left(t_s+\frac{D}{c}\right)-\Phi_0-\frac{\pi}{4}+\frac{3}{4}\left(\frac{G\,M_c^{\text{eff}}}{c^3}\hspace{1mm}8\pi f\right)^{-5/3}+\arg (1-Be^{\mp i2\varphi})\\
    &\approx 2\pi f\left(t_s+\frac{D}{c}\right)-\Phi_0-\frac{\pi}{4}+\frac{3}{4}\left(\frac{G\,M_c^{\text{eff}}}{c^3}\hspace{1mm}8\pi f\right)^{-5/3}\pm \frac{B\sin(2\varphi)}{1-B\cos(2\varphi)},
\end{aligned}    
    \end{equation}
where in the last line $B\ll 1$ was assumed.

\section{Numerical analysis\label{sec:validity}}
Throughout our analysis, we have worked under the assumption that the presence of the third small body does not affect the equations for the Keplerian orbits of the binary system, while the GW emission and the resulting backreaction have been estimated under the semi-Keplerian approximation. Since the deviations from the binary expressions will be most significant when $m_3/\mu$ is large (see Fig.~\ref{fig:g_function}), the important question to address here is: How big can $m_3$ be compared to $\mu$ before the assumptions of the CR3BP break down? For that purpose some \text\tt{Mathematica} code has been developed in which the classical Three-Body-Problem has been implemented, albeit without any relativistic effects or GW emission.

The first important assumption is that the binary's orbit in the presence of the third body keeps being circular, in other words, that the eccentricity of the orbit is $e=0$. To test this assumption we have kept track of the eccentricity of the binary by computing the Laplace-Runge-Lenz (LRL) vector $\Vec{A}$ associated to the binary separation $\Vec{r}$ along the trajectory, which is defined as
\begin{equation}
    \Vec{A}=\Vec{p}\times \Vec{L}-G\,M\,\mu\hspace{1mm}\frac{\Vec{r}}{||\Vec{r}||}\hspace{1mm}.
\end{equation}
It can be shown\footnote{See Ref.~\cite[p.91]{celestial}.} that the eccentricity of an orbit described by an object of mass $\mu$ under the gravitational influence of an object of mass $M$ is related to the LRL vector via
\begin{equation}
    e=\frac{||\Vec{A}||}{G\,M\,\mu}\hspace{1mm}.
\end{equation}

\begin{figure}[!t]
    \centering
    \includegraphics[scale=0.7]{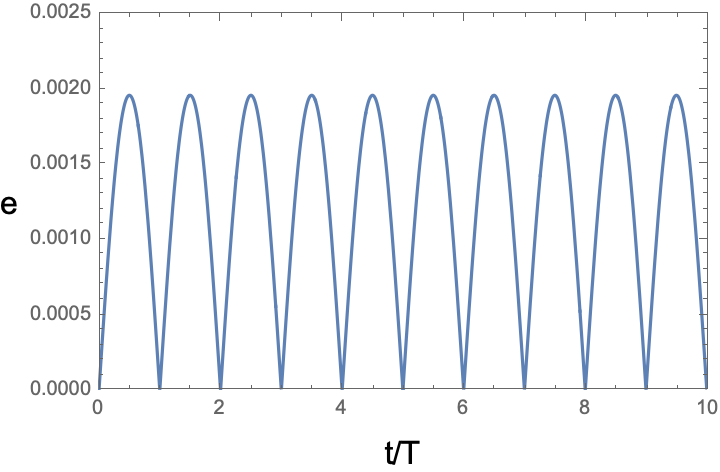}
    \caption{Change in the eccentricity of a binary with mass ratio $\eta=0.02$ in the presence of a co-rotating object with mass $m_3=0.05\mu$, through 10 orbits of the system. $T$ is the keplerian period.}
    \label{fig:eccentricity_evolution}
\end{figure}

For the two body case, the LRL vector is a constant of motion, and so is the eccentricity, as expected. By introducing the third body's interaction, the LRL vector associated to the binary separation is no longer guaranteed to be constant, and so the eccentricity of the binary orbit can change through time. In Fig.~\ref{fig:eccentricity_evolution} we show the evolution of the eccentricity for an instance with $\eta=0.02$ and $m_3/\mu=0.05$ .

\begin{figure}[!t]
    \centering
    \includegraphics[scale=0.52]{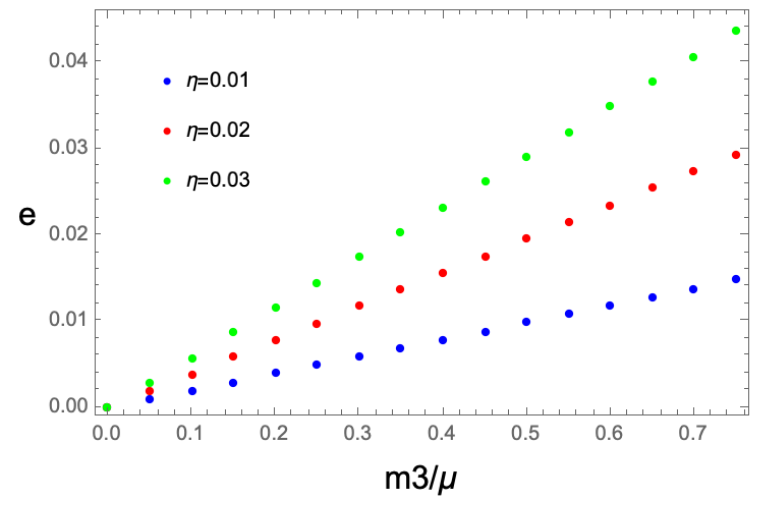}
    \caption{Maximum eccentricity achieved by the parent binary after 100 orbits in terms of $m_3/\mu$ for different mass ratios $\eta$.}
    \label{fig:eccentricity_results}
\end{figure}

In order to quantify the deformation, we have set the criterion of taking the maximum value of $e(t)$ over several periods as the reference value, which represents the maximum deviation from circularity throughout the evolution of the system. From the previous plots we observe that even for masses as big as $m_3/\mu=0.05$ the eccentricity of the binary stays below $e\leq 0.002$, meaning that the circular orbit approximation is quite well preserved. The same calculation has been repeated for several other values of $\eta$ and $m_3/\mu$ (for this last parameter, always keeping $\mu$ constant), as shown in Fig.~\ref{fig:eccentricity_results}.

\begin{figure}[!t]
    \centering
    \includegraphics[scale=0.45]{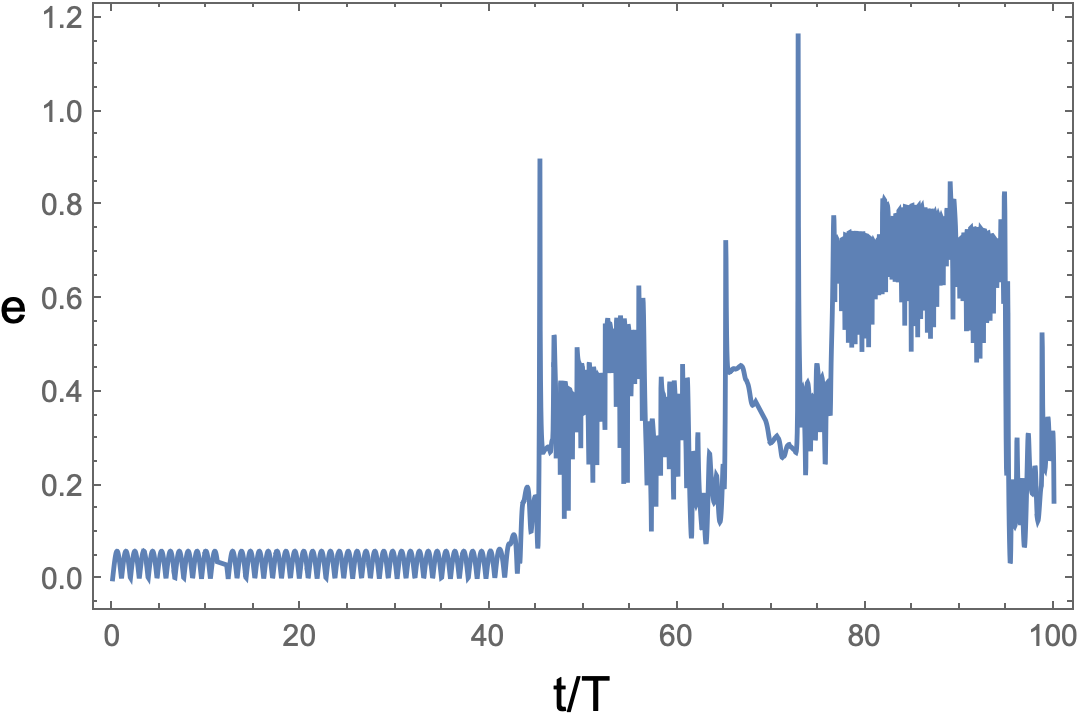}
    \caption{Evolution of the binary's eccentricity along 100 orbits for an instance with $\eta=\eta_\mathrm{crit}$ and $m_3=0.8\mu$. After approximately 40 periods the parent binary ceases to follow circular orbits and the system becomes chaotic.}
    \label{fig:blowup}
\end{figure}

As expected, the eccentricity grows as $m_3/\mu$ grows, but in a rather controlled manner as long as sensible ranges are chosen. The other observation drawn from Fig.~\ref{fig:eccentricity_results} is that the more asymmetric the parent binary is, the smaller the effect of the third body is in its eccentricity. In Fig.~\ref{fig:blowup}, we show a case that pushes the parameters beyond sensible limits, resulting in a chaotic behavior of the system in the long run.

Measuring the evolution of this eccentricity $e$ as the system evolves provides a measure of the deviation from circularity of the parent binary's orbit. However, only by itself it does not fully characterize the extent to which the assumptions that support the theoretical formalism are violated, as we show in the following paragraph.

As soon as the third body has some mass and velocity, the total momentum of the whole system in the binary's CM frame is not null, which means that the CM of the system will slowly ``drift away" from the chosen reference frame. Moving to the whole system's CM frame would not exactly solve the issue, as the equations for the binary's orbit in this new frame would slightly differ from the original ones, upon which all the well known results regarding the binary system's GW emission are based.

\begin{figure}[!t]
    \centering
    \includegraphics[scale=0.36]{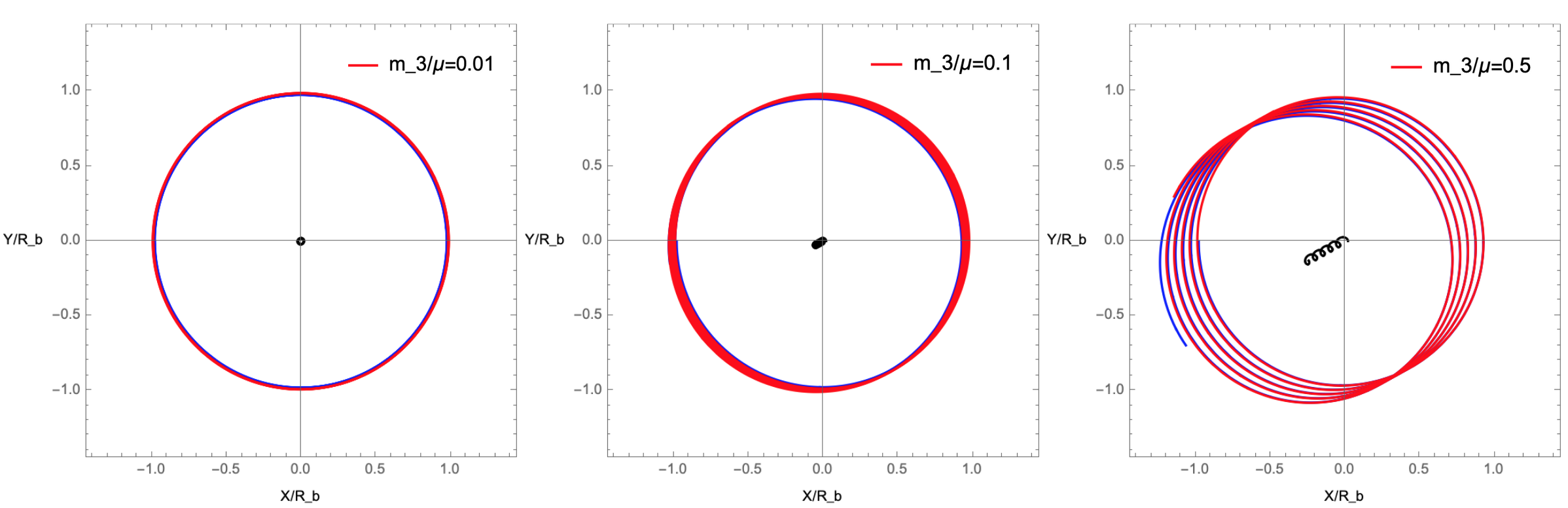}
    \caption{5 orbits of the CR3BP system for various values of $m_3/\mu$.}
    \label{fig:drift}
\end{figure}

If this ``drift velocity" between the reference frames is small enough it could be regarded as a peculiar velocity of the source itself, which in the context of the study of coalescing binaries at cosmological distances is usually ignored. In Fig.~\ref{fig:drift}, we show the aforementioned ``drift" of the system for various values of the $m_3/\mu$ parameter.

\begin{figure}[!t]
    \centering
    \includegraphics[scale=0.3]{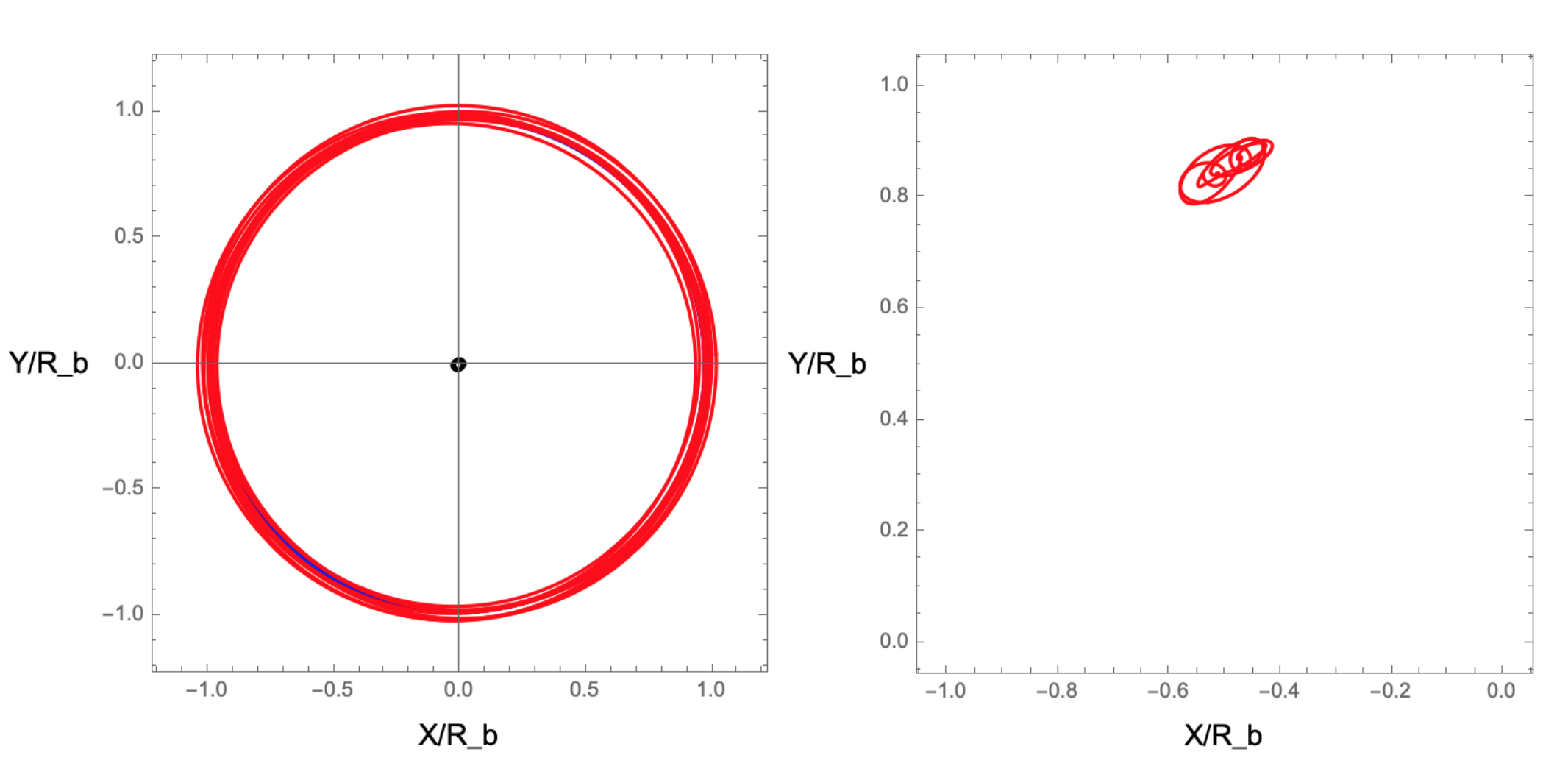}
    \caption{Left: Orbit of the third body for 10 orbits when initial position is disturbed by 2\%. Right: Same as the left panel but in the co-rotating frame (unperturbed initial position would correspond to a fixed point).}
    \label{fig:oscil}
\end{figure}

One final approximation that might not be too realistic and that could be the key to hunt down the presence of the co-rotating object in experimental data is the assumption that the third body always sits exactly at the $L_4$ or $L_5$ Lagrange point though the evolution of the system. As already mentioned, it can be shown that when perturbed the third body oscillates around the Lagrange points with frequencies given in Eq.~\eqref{epicycles}. We demonstrate that behavior in Fig.~\ref{fig:oscil}.

Accounting for these oscillations would require correcting the orbit equations of Eq.~\eqref{3trajectory}. Even if the amplitudes of the epicycles were small, the presence of frequencies other than $\omega_s$ (namely $\omega_{1,2}$) in the orbit equations would probably result in interesting modifications to the results presented so far, specially in the quantities concerning the frequency domain.

\section{Discussion \label{sec:discussion}}

The uncertainty introduced in the determination of the chirp mass (which can be of the order of a few percent, see Fig.~\ref{fig:chirpmass}), is still negligible compared to the typical error bars found in the literature~\cite{94events,03b,population}. Moreover, it is important to remark that not all compact binary systems are suitable to trap a 1:1 resonating object. In that sense, the stability condition imposed by Eq.~\eqref{etacrit} is probably one of the harshest constraints on the problem. The need for this high asymmetry heavily restricts the part of the merging compact binary population that could potentially be a host to a resonating third small body. From the detection event candidate list available in the literature, GW191219$\textunderscore$163120 is the one with lowest mass ratio to date~\cite{our_candidate}, with an estimated value of $\eta\approx 0.036<\eta_\mathrm{crit}$. This event is usually not present in the different catalogs found in the literature~\cite{94events}, given that its False Alarm Rate (FAR) or other detection parameters do not pass the imposed thresholds.

However, as it is pointed out in Refs.~\cite{03b,our_candidate}, events with mass parameters outside of the typical ranges used for calibrating template waveforms may introduce considerable uncertainties in their $p_\mathrm{astro}$. This fact acknowledges the need to update the waveform filtering and data analysis section of the signal processing part and review the systematic errors that come into play when considering less typical regions of the binary's population's parameter space, such as the mass ratio $\eta$. The number of events detected to date is probably not enough for meaningful statistics of the asymmetry distribution~\cite{population} and potential biases of current ground based interferometry detectors towards more symmetric binaries might be significant. 

Finally, if the third body is assumed to be a NS, its mass needs to be at least around the Chandrasekhar threshold mass of $m_3\sim 1.4M_{\odot}$, which due to the asymmetry constraints leads to the conclusion that $m_1\gtrsim 10^3M_{\odot}$, a BH mass range that current detectors have not been able to access so far, but that further observation runs of LVK detectors~\cite{Guo:2022sdd} or the third generation detectors like ET~\cite{Huerta:2010tp} can in principle probe. 
Moreover, if we consider the scenario in which the binary is formed by two supermassive black holes~\cite{Jiang:2022aek} or the combination of a supermassive and an intermediate-mass black holes~\cite{Gair:2010yu}, then the third body could be a black hole with a mass anywhere in the range $m_3\in[1.4,100]M_{\odot}$. In light of Eq. \eqref{new_frequencyy}, the frequency range at which the CR3BP emits is not substantially modified as long as the mass of the third body remains within its range of validity, since the effective chirp mass does not differ significantly from that of the binary. Thus, for either of the scenarios described, future space-based missions like LISA would be sensitive to these triple systems when the binary's mass falls within their detectability range~\cite{Gair:2017ynp,Colpi:2024xhw}. However, whether the expected number of detection events that fall within the allowed asymmetry range for the CR3BP will be significant or not is still an open question.

Another significant limiting factor for the applicability of the results presented in this paper is that by working in a semi-Keplerian setting and not accounting for PN corrections the validity of the results is restricted to the rather-far end of the inspiralling phase. Given that current detectors can only access the last minutes/seconds before the merger, it will not be until future detectors register the earlier stages of the inspiralling phase that the CR3BP formalism gains importance.

All in all, serious applicability of the formalism of GWs in the CR3BP will need to wait for both further analytical and theoretical refinements as well as the development of future third and fourth generation GW detectors that will have the capabilities to detect signals from scenarios where our theory applies. The results presented in this paper are just the basis for the meticulous study of GWs in the CR3BP. They provide the first order corrections to the binary results, but there are numerous qualitative improvements to be done.

In the first place, accounting for 1PN and probably higher order corrections is paramount if any serious search in experimental data is to be considered in the future. 1PN corrections to the equations of motion in the CR3BP have been known for quite some time now~\cite{PNcorrections}, but they have not been used for GW considerations. Among other quantities, the Lagrange positions (see also Ref.~\cite{Schnittman:2010br}), Kepler's Third Law or the critical mass ratio are altered by these PN corrections:
\begin{equation*}
     X_L=R_b\left(\eta-\frac{1}{2}\right)\left(1+\frac{5}{8}\frac{R_s}{R_b}\right),\hspace{1cm}Y_L=\frac{\sqrt{3}}{2}R_b\left(1+\frac{(6\eta\,(1-\eta)-5)}{24}\frac{R_s}{R_b}\right),
\end{equation*}
\begin{equation}\label{PNcorrections}
    \omega_s = \sqrt{\frac{G\,M}{R_b^3}}\left(1+\frac{(\mu/M-3)}{4}\frac{R_s}{R_b}\right),\hspace{1cm}\eta_\mathrm{crit}= 0.03852-0.14528\hspace{1mm} \frac{R_s}{R_b}\hspace{1mm},
\end{equation}
\phantom{a}\\
where $R_s$ is the Schwarzschild radius associated to the mass $M$. Including these corrections would certainly complicate the analytical forms of the expressions presented in Section 3, but as it is obvious from Eq.~\eqref{PNcorrections}, they become relevant when $R_b$ is comparable to $R_s$, i.e. when approaching the merging phase.

The correction to the critical mass ratio in Eq.~\eqref{PNcorrections} already hints a fact that is quite intuitive: as the system evolves and the objects come closer and they gain velocity, it is expected that the third body will eventually be expelled from the system. The ejection of resonating third bodies has already been considered in Ref.~\cite{3bodyastro}, where 1PN corrections and GW dynamics were taken into account but all computations were performed numerically. One of the conclusions of the paper was that the ejected body could have velocities of up to $0.14c$, which should in principle leave a very characteristic signature in the GW emission spectrum and could be looked for in the data. In particular, Ref.~\cite[Fig.3]{3bodyastro} illustrates this ejection scenario.

In the spirit of continuing with the analytical development of GWs in the CR3BP, an interesting next step that resonates more with the kind of results presented in this paper could be to explore the effects of adding the small quasi-periodic oscillations around the equilibrium point in the equations of motion. The starting point would be to modify Eq.~\eqref{3trajectory} in the following way:
\begin{equation}\label{epitraj}
\begin{cases}
    x(t)=R_L\cos(\omega_s t+\pi+\varphi)+R_1\cos[(\omega_s\pm\omega_1)t+\varphi_1]+R_2\cos[(\omega_s\pm\omega_2)t+\varphi_2],\\ y(t)=R_L\sin(\omega_s t+\pi+\varphi)+R_3\sin[(\omega_s\pm\omega_1)t+\varphi_3]+R_4\sin[(\omega_s\pm\omega_2)t+\varphi_4], \\z(t)=0,
\end{cases}
\end{equation}
where $R_1,...,R_4$ would need to be small compared to $R_L$, the angular frequencies $\omega_1,\omega_2$ are given in Eq.~\eqref{epicycles} and the phases $\varphi_1,...,\varphi_4$, are chosen at random (in reality they would depend on the initial conditions of the perturbed particle). This model for the perturbed orbit is motivated by the discussion in Ref.~\cite[p.176]{celestial}. Rather than the exact effect that these small oscillations would have in the emitted power and similar quantities, it would be interesting to see how the presence of angular frequencies other than $\omega_s$ would affect quantities like $\Psi_+,\Psi_\times$, which are important quantities for the detection of signals among the noise. This approach is currently being developed by the authors.

To finish up, a proper validation of the analytical results demands comparison against numerical codes that implement the full back-reaction effect of GWs, which would require some slight modifications in the numerical codes.

\section{Conclusions \label{sec:conclusions}}

In this paper, we have studied the GW emission during the inspiralling phase in the Circular Restricted Three Body Problem.

Starting from a semi-Keplerian treatment of the problem and laying out the necessary constraints and approximations for the relevant parameters, expressions for the polarization amplitudes of the GWs and the associated emitted power have been calculated, which are given by Eq.~\eqref{polarizations3} and Eq.~\eqref{3power}, respectively. These results can be seen as corrections to the analogous expressions for the binary case, the magnitudes of which depend on the key parameter $m_3/\mu$.

Accounting for the orbital energy loss due to GW emission has led to Eq.~\eqref{new_freq}, which is one of the crucial results. This equation can be interpreted as an effective rescaling of the binary's chirp mass, defined in Eq.~\eqref{newchirp}, practically introducing a new uncertainty in the determination of this quantity from experimental data. In physical terms, one of the consequences of Eq.~\eqref{new_freq} is that the presence of the third body actually slows down the coalescence of the system, as suggested by  Eq.~\eqref{time_to_coalescence}.

The effect of the resonating small third body cannot be fully replaced by the introduction of this new effective chirp mass, as it is shown via the computation of the Fourier transforms of the polarization tensors in Eq.~\eqref{3fourier} and energy emission during the inspiralling phase in Eq.~\eqref{acabateya}. This last expression also shows that, even if the GW emission power decreases, the total energy emitted is greater than in the only-binary scenario.

Finally, in Sec.~\ref{sec:validity}, some numerical results are shown to assess the range of values of $m_3/\mu$ for which the CR3BP assumptions remain sensible (Figs.~\ref{fig:eccentricity_results} and \ref{fig:drift}). Although a threshold value for the $m_3/\mu$ parameter is not provided, simulations where $m_3/\mu<0.05$ provide stable and satisfactory results after more than 500 periods of the system. To provide a better justified threshold value, apart from having to sort out a certain element of arbitrariness, numerical codes that include all the relativistic effects need to be implemented.

To sum up, the CR3BP is a peculiar scenario in GW astronomy that is likely to gain interest as the number, quality and variety of events detected by current and future detectors is set to increase in the near future. Building on the work presented in this paper and overcoming its limitations, for instance by considering PN corrections and small oscillations in the trajectories, the first GW detection of a three body system is a milestone that could be reached in the not so distant future.

\section*{Acknowledgements}
The authors acknowledge support from the research project PID2021-123012NB-C43 and the Spanish Research Agency (Agencia
Estatal de Investigaci\'on) through the Grant IFT Centro de Excelencia Severo Ochoa No CEX2020-001007-S, funded by MCIN/AEI/10.13039/501100011033. S.K. is supported by the Spanish Atracci\'on de Talento contract no. 2019-T1/TIC-13177 granted by Comunidad de Madrid, the I+D grant PID2020-118159GA-C42 funded by MCIN/AEI/10.13039/501100011033, the i-LINK 2021 grant LINKA20416 of CSIC, and Japan Society for the Promotion of Science (JSPS) KAKENHI Grant no. 20H01899, 20H05853, and 23H00110. M.M. acknowledges support from the JAE Intro ICU grant no. JAEIntroICU-2022-IFT-14.

\section*{References}
\bibliographystyle{iopart-num}
\bibliography{references}

\end{document}